\documentclass{aa}
\begin{document}

%   \thesaurus{01     % A&A Section 1: Letters 
%              (10.06.1;  % Galaxy: formation,
%               10.07.2;  % Galaxy: globular clusters: general,
%               10.08.1;  % Galaxy: halo,
%               10.11.1;  % Galaxy: kinematics and dynamics,
%               10.19.3)} % Galaxy: structure.
%
   \title{Structure in the velocity space of globular clusters}

%   \subtitle{The falling down globular clusters}

   \author{E.~J. Alfaro\inst{1}
          \and A.~J. Delgado\inst{1}   
          \and M.~A. G\'omez-Flechoso\inst{2}
	  \and F. Ferrini\inst{3}
	  \thanks{\emph{On leave of
              absence from:} Universit\'a di Pisa, Piazza Torricelli 2, 56100 
              Pisa, Italy}
          \and I. Castro\inst{1}		
%\fnmsep\thanks{Just to show the usage
%          of the elements in the author field}
          }

   \offprints{E.~J. Alfaro, \email{emilio@iaa.es}}

   \institute{Instituto de Astrof\'{\i}sica de Andaluc\'{\i}a (CSIC),
              Apdo. 3004, Granada 18080, Spain
             \and
             Observatoire de Gen\`eve, CH-1290 Sauverny, Switzerland
	      \and 
             INTAS, 58 Avenue des Arts, 1000 Bruxelles
		}

   \date{Received          / Accepted            }

   \abstract{
     We present an analysis of the velocity space of a sample of
     globular clusters (GC) with absolute proper motions. The vertical
     component of the velocity is found to be correlated with
     luminosity and galactocentric radius. We divided the sample into
     two luminosity groups above and below the peak of the luminosity
     function (LF), M$_\mathrm{V}$=$-$7.5, for Galactic GCs. The two groups
     display different kinematic behaviour according to the first and
     second statitical moments of the velocity distribution as well as
     distinct velocity ellipsoids. The velocity ellipsoid of the high
     luminosity clusters is aligned with the symmetry axes of the
     Galaxy, whereas the minor axis of the Low Luminosity group is
     strongly inclined relative to the Galactic rotation axis.
   }

  \maketitle
%
%________________________________________________________________

\section{Introduction}

It has been known since the late eighties that  groupings exist in the velocity
space of several halo tracers which could be interpreted as debris from larger
stellar sub-structures disrupted by the Galaxy (Sommer--Larsen \& Christensen
1987, Dionidas \& Beers 1989, 
Arnold \& Gilmore 1992, Poveda et al. 1992).
In particular, some kinematic studies of halo stars (Majewski et al. 1996, 
Chen 1998) indicate that the Galactic halo may not be a dynamically relaxed 
system. The presence of three moving groups in the SA57 field near the NGP,
with different metallicity distributions, supports the hypothesis that the
Galactic halo is mainly formed from a mixture of several stellar streams.

Analysis of the radial velocity dispersion tensor of the GCs (Hartwick
1996)  also falls upon the idea that the Galactic halo is
not dynamically homogeneous. 
Two different subsystems are clearly identified in that
study; one, located in the outer region of the halo (R$_\mathrm{g}>$ 7
kpc), shows a minor axis parallel to the Galactic rotation axis, 
while the second inner one is highly inclined relative to the
symmetry axes of the Galactic disk. The fact that the inner Galactic
GCs  present a velocity ellipsoid almost parallel to the
spatial distribution of the outer satellites suggests that the
outer satellites may be outlining the Galaxy's dark matter halo and that
the actual residual velocity distribution of the inner halo clusters 
might  be representative of the dominant potential well in the early phases
of the halo formation (Hartwick 2000).  Thus, there is evidence in favor 
of two peculiar kinematic features in the halo: 1) miss-aligned residual
velocity ellipsoids and 2) moving groups whose origin might be ascribed  either
to  ``pollution'' by disrupted satellites or to the signature of the early 
dominant potential well.   

Another interesting peculiarity of the Galactic globular cluster system is
its present  LF. In contrast to disk open clusters that show a
monotonically increasing LF, globular clusters show a peaked distribution
with a maximum around M$_\mathrm{V}\,\approx\,-$7.5.  Some authors consider  
this LF  to be primordial (e.g. Fall \& Rees 1988, Fritze-von Alvensleben 1999) while the most accepted interpretation supposes that the present distribution
evolved from an initial power-law distribution through  destructive  processes 
(Larson 1996, Elmegreen \& Efremov 1997). Dynamical modelling of halo 
globular clusters in the Milky Way potential  shows that destructive
processes, and their time scales, depend strongly on orbital parameters and
cluster masses  (e.g. Capuzzo-Dolcetta 1993). 

Following these arguments, we ask  to what extent kinematics and
luminosity  (and hence mass) are correlated and, if correlated, how this
reflects  on the velocity space of the GCs. In this respect, the work by
Burkert \& Smith (1997)  indicates that the metal-rich GCs can be separated
by mass into three groups, with different spatial and kinematic properties. 
Here we present a similar analysis for the metal-poor GCs and study the
velocity space of the halo GCs with complete kinematic information.

The paper is organized into three main sections. Section two is devoted
to a description of the sample and presentation of our results while the final
section considers possible interpretations.

\section{Kinematics and Luminosity}

%%
%%                                                One column figure
%%----------------------------------------------------------- S_vib
\begin{figure}
\vspace{12cm}
\includegraphics{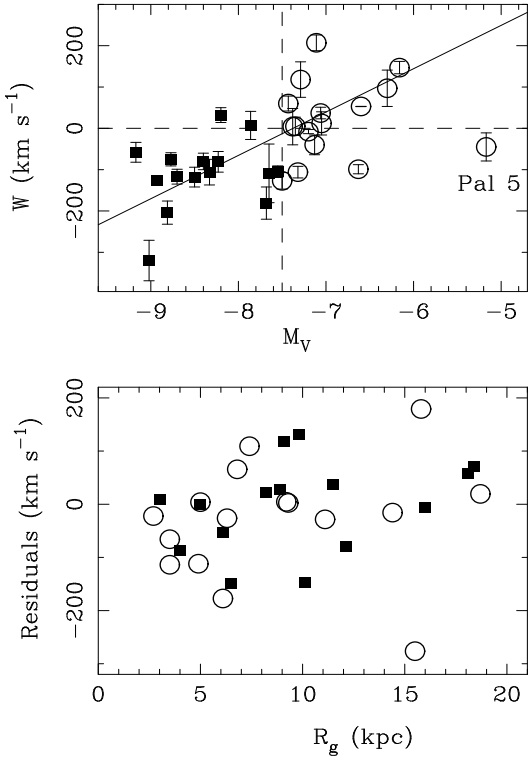}
%%\rule{0.4pt}{4cm}% line thickness, height of picture
\vskip -1.3truecm
\caption{({\it Top panel}). Vertical component of the velocity versus M$_\mathrm{V}$
for the sample of halo GCs with complete kinematic information. Open
circles represent the clusters with M$_\mathrm{V}\geq-$7.5 and black squares those
brighter than M$_\mathrm{V}$=$-$7.5. Note that most of the clusters appears located 
in two of the four quadrants in which has been divided the plot. ({\it Bottom
panel}). Residuals of the linear fit shown in the top panel versus
galactocentric distance.}
\label{fig1}
\end{figure}
%
%______________________________________________________________

\subsection{The sample}

Our data sources are the compilation of absolute proper motions for GCs
(Dinescu et al. 1999; DGA in the following) and the updated version 
(June 1999) of the catalogue  
of ``Milky Way Globular Cluster Parameters'' 
by Harris (1996). The first
compilation provides information about the velocity components and orbital
parameters, as well as metallicity, radial velocity and spatial information
for 38 Galactic GCs (the largest sample so far  with complete kinematic
information). The second one provides a large set of physical, 
structural  and photometric parameters,  including total luminosity,
for the entire Galactic globular cluster system. Errors in velocities  have
also been taken from DGA, who adopted a 10\% error in 
the distances. The average uncertainty in the integrated absolute magnitudes is
$\approx$0.5. 
     
We limit our study to the halo GCs within a galactocentric radius of 20 kpc. 
Thus, two clusters with [Fe/H]$\geq-$0.9, typical of the disk sub-system, 
and \object{Pal~3} located well beyond our limit radius, have been removed. 
Three other clusters (\object{NGC~6254}, \object{NGC~6626}, and
\object{NGC~6752}) display disk-like 
orbits and can be considered to be the metal-poor tail of the disk sub-system
(DGA). \object{NGC~5139}, in addition, is thought to be the core of a
disrupted  dwarf spheroidal galaxy 
(Majewski et al. 2000). These four clusters
are consequently  omitted from our analysis. The final kinematic sample
contains 31  ``bona fide'' halo GCs representing $\approx$40\% of the  halo
cluster population within 20 kpc. This sample distributes with metallicity,
total luminosity, and galactocentric radius in a  rather similar way to the
ones shown for the  halo  cluster population inside  the same volume. 
The main difference involves the R$_\mathrm{g}$ variable whose distribution is
flatter than the typical potential law shown by the halo GCs.

The kinematic data are considered in a cylindrical coordinate system. The
$\Pi$  component is positive outwards from the galactic center, whereas the
other components retain their usual conventions; $\Theta$ is positive
towards the direction of galactic rotation and W towards the North
galactic pole. 
The centroid and dispersion of the velocity components for our sample
[($\langle\Pi\rangle$=25$\pm$25, $\langle\Theta\rangle$=50$\pm$16,
$\langle$W$\rangle$=$-$43$\pm$21); ($\sigma_\Pi$=137$\pm$23,
$\sigma_\Theta$=100$\pm$13, $\sigma_\mathrm{W}$=107$\pm$21)] present a mean rotational
value similar to the ones obtained from radial velocity data for the
metal-poor 
globular clusters (C\^ot\'e 1999). The present values of the velocity 
dispersion  are
in  good agreement with those obtained by other authors for halo stellar
samples (Norris 1986, Morrison et al. 1990). Thus, on the basis of rotation 
and velocity dispersions, the sample
can be considered as  representative of the halo globular clusters. 

%                                     Two column figure (place early!)
%______________________________________________ Gamma_1 (lg rho, lg e)

\begin{figure*}
\vspace{5.5cm}
\includegraphics{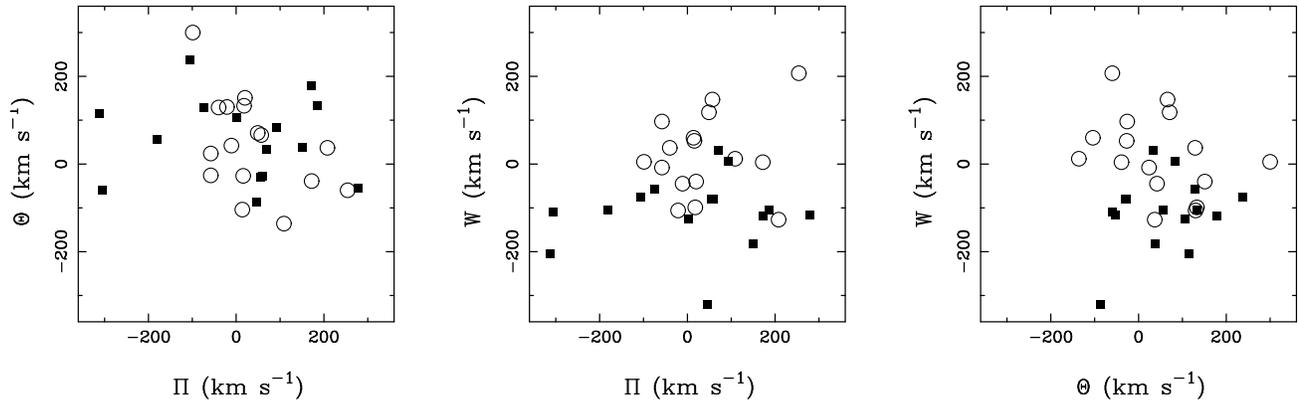}
%\hfill     
\parbox[]{17.5cm}{\caption[]{Distribution of the  halo globular clusters with 
complete kinematic information  onto the  main planes defined by the
cylindrical velocity components. Symbols as in fig. 1.

\label{fig2}
}}

\end{figure*}
%
%__________________________________________________________________

\subsection{Velocity components versus Luminosity}

We began our study by analyzing the distribution of the velocity components
with luminosity for our sample. The vertical component (W) is plotted versus
the integrated absolute magnitude (M$_\mathrm{V}$) in fig. 1 (top
panel). A clear correlation is apparent in this plot,
which translates into a probability lower than 0.1\% that the two variables are
uncorrelated  ($\tau$-Kendall and Spearman tests). Only the lowest luminosity
cluster in our sample, Pal 5, separates from the main distribution. This cluster 
is  representative of a small group of faint objects located beyond 
R$_\mathrm{g}$=10
kpc which do not have a counterpart in the inner galactic regions (McLaughlin
2000). It has been suggested that this group of clusters formed more recently
than the rest  (van den Bergh 1999, private communication to
MacLaughlin). Fig. 1 (top panel) also  shows a robust linear fit to
the data while the residuals of the fitting are plotted against
galactocentric radius in the bottom panel of fig. 1, where a weaker but
apparent correlation (probability lower than 7\% that the two variables are
not correlated according to $\tau$-Kendall and Spearman tests) is
also present. These results indicate that for halo clusters with
integrated absolute magnitudes between $-$9.2 and $-$6.0, the vertical
component of the velocity scales with  luminosity and, marginally, 
with galactocentric radius. 

The other velocity components do not clearly   correlate  with
luminosity although, as we will discuss in the next sub-section,
clusters with different luminosity display distinct kinematic behaviours.

\subsection{Luminosity Groups}

Our sample has been divided into two groups according to their
integrated absolute magnitude. There are 15 clusters brighter than
M$_\mathrm{V}$=$-$7.5, which form the {\sl High Luminosity Group} (HL) while
16 constitute the {\sl Low Luminosity Group} (LL). The
first and second moments of the velocity distribution have
been estimated for both groups and the mean radial component shows a
marginal difference between the HL and LL clusters (9$\pm$40 and 39$\pm$24
respectively), where the LL group shows evidence for a weak expansion.  
The mean rotational moment (57$\pm$24 and 43$\pm$28 for HL and LL
respectively) is similar for both groups and also agrees with the
average value obtained for the metal-poor GCs from radial velocity data
(C\^ot\'e 1999).

The main difference involves the vertical
velocity component where 13/15 of the HL clusters display negative
values of W and  an  average vertical component of $-$109$\pm$21. In
contrast, the mean value for the LL group is 20$\pm$23, where
10/16 objects show positive W values. A Kolmogorov-Smirnov two-sample test
gives a probability lower than 1\% that both sub-samples come from the same
population.

Fig. 2 shows the velocity space for our data, projected onto the three 
principal planes defined by the cylindrical velocity components.  This plot
reveals that the luminosity groups distribute in a
different way: 1) the radial component of the HL clusters (black squares)
display a wider range of values than the LL group (open circles) and 2) the 
 ($\Pi$, W) plane distribution of the LL clusters is
highly inclined relative  to the Galactic rotation axis. The velocity
dispersions for the LL and HL clusters are ($\sigma_\Pi$=168$\pm$32,
$\sigma_\Theta$=92$\pm$15, $\sigma_\mathrm{W}$=81$\pm$17) and   
($\sigma_\Pi$=97$\pm$19,
$\sigma_\Theta$=107$\pm$21, $\sigma_\mathrm{W}$=90$\pm$17) respectively.

In order to analyze  in more detail the velocity space of these groups, we
have derived the velocity ellipsoid for both distributions. The
evaluation procedure  has been configured in a bootstrap loop in order to
provide an  estimate of the parameter uncertainties. Table 1
shows the module and  direction of the three main axes of the two
distributions. The velocity ellipsoid defined by the distribution of the 
brightest clusters is almost  parallel to the principal axes of the Galaxy, 
while the Low Luminosity objects distribute in a highly inclined
ellipsoid.    

%__________________________________________________ One column table
   \begin{table}
      \caption[]{Velocity ellipsoids for the two groups of luminosity,
      $\sigma$ is given in km~s$^{-1}$ and the Galactic coordinates,
      defining the direction of the main axes, in degree. The estimated
      errors are the dispersion of 100  bootstraped samples.}
         \label{KapSou}
      \[
         \begin{array}{r r r r r r}
            \hline
            \noalign{\smallskip}
      &     HL      &          &            &    LL      &            \\
            \noalign{\smallskip}
            \hline
            \noalign{\smallskip}
  \sigma~~~~ &      l ~~~~ &   b ~~~~ & \sigma~~~~ &     l~~~~  &     b~~~~  \\
            \hline
            \noalign{\smallskip}
    151 \pm 21 &   10 \pm 40 &  ~6 \pm ~9 & 105 \pm 14 & -63 \pm 50 &  45 \pm 32 \\
    130 \pm 21 &  -80 \pm 42 &  ~4 \pm 26 & 102 \pm 14 &  40 \pm 40  & 13 \pm 31 \\
     80 \pm 15 & -137 \pm 58 &  83 \pm 29 &  50 \pm 11 & -38 \pm 24 & -42 \pm 43 \\
            \noalign{\smallskip} 
            \hline

         \end{array}
      \]
   \end{table}
%
%_________________________________________________________________________

\section{Discussion and Conclusions}

The analysis performed in the previous section provides evidence for  
a clear connection between kinematics and luminosity which can be   
summarized as follows:

   \begin{enumerate}
      \item The vertical component of the velocity (W) scales with
      luminosity and galactocentric radius for a sample of 31 metal-poor 
      GCs with absolute proper motions.
         
      \item  The sample of globular clusters has
       been divided into  two absolute magnitude groups separated above and
        below M$_\mathrm{V}$=$-$7.5. The first and second statistical moments of  the
        velocity distributions of the groups show significant differences.        

      \item Both luminosity groups display different structure in the
       velocity space. The velocity ellipsoid of the High Luminosity group
      is aligned with the main axes of the Galactic
       disk while the Low Luminosity clusters show a velocity
          ellipsoid  with  minor axis highly inclined with respect to the
Galactic rotation  axis.

   \end{enumerate}
 
The first result is very surprising and we have no satisfactory explanation
for the existence of such a relationship between vertical velocity and
luminosity. 
Given that our kinematic sample  represents  40\% of the halo
population in this volume, we consider whether this correlation could be
produced by a selection effect involving the clusters with absolute proper
motions. In order to check this we have taken the complete
set of halo GCs within  20 kpc and estimated the average vertical component
for the two luminosity groups using  only radial velocity data.
Assuming the system has neither net expansion nor rotation, we obtain ($\langle
\mathrm{W}\rangle^\mathrm{HL}$=$-$32$\pm$6; $\sigma^\mathrm{HL}_\mathrm{los}$=115) and  ($\langle
\mathrm{W}\rangle^\mathrm{LL}$=46$\pm$7; $\sigma^\mathrm{LL}_\mathrm{los}$=128) for the HL and LL groups
respectively.  After correcting
for rotation, assuming $\langle\Theta\rangle$=50 km~s$^{-1}$,  we obtain
($\langle\mathrm{W}\rangle^\mathrm{HL}$=$-$45$\pm$6; $\sigma^\mathrm{HL}_\mathrm{los}$=112)
and  ($\langle\mathrm{W}\rangle^\mathrm{LL}$=55$\pm$7; 
$\sigma^\mathrm{LL}_\mathrm{los}$=128).
These results go in the direction marked in fig. 1, clusters with low
luminosity move, onn average,  towards the NGP while clusters brighter
than M$_\mathrm{V}$=$-$7.5 show a negative mean W component.

In addition the second and third items noted above clearly show that both
luminosity groups occupy different volumes in the velocity space. They 
can be  distinguished by the centroids of the distributions as well as by the
main axes of the dispersion ellipsoids. As noted above, Hartwick (1996) pointed out 
that the metal-poor GCs display different dispersion tensors for objects located
within and beyond the solar galactocentric radius. The inner clusters show
a dispersion tensor highly inclined with respect to the Galactic rotation
axis while the outer group is almost aligned with  the Galactic symmetry
axes. Could our results and the results of Hartwick represent different
aspects of the same phenomenon?. If this hypothesis is correct then luminosity
and galactocentric radius should correlate in the sense that Low Luminosity
clusters should be preferentially located in the inner Galactic
regions. However, this does not appear to be the case. Our sample does not
show any correlation between integrated absolute magnitude and
galactocentric radius and we extend this conclusion to the
entire  Galactic system  of GCs (McLaughlin 2000). Therefore  the connection
between kinematics and luminosity stressed in this work can not be
accounted for only by different episodes of cluster formation in distinct
gravitational potentials. We must devise another mechanism that is mainly
driven by luminosity.

McLaughlin (2000) recently showed that globular clusters fit a plane
in the parameter space defined by  binding energy, luminosity and
galactocentric radius. It was suggested that this relationship is primordial. 
Our correlation between the vertical velocity component of the GCs 
and the same two variables (luminosity and galactocentric radius), gives rise to the 
possibility that  this correlation
is also primordial or that it reflects destructive processes that have occurred
since the halo formation. We do not have a definite answer to this question,
but our analysis suggests that ``external'' variables, such as the orbital 
parameters (location in the velocity space) are intimately connected with 
``internal'' parameters such as the binding energy and/or luminosity.

\begin{acknowledgements} We are very grateful to Antx\'on Alberdi, Jos\'e
Franco and Enrique P\'erez for useful discussions. Jack Sulentic is acknowledged for
his careful revision of the language. This work 
has been partially supported by the Spanish DGICYT, through grant
PB97-1438-C02-02 and by the Research and Education Council of the 
Autonomous Government of Andaluc\'\i a (Spain). Spanish CICYT under grant 
ESP98-1339-C02-02 has partially funded this work.
\end{acknowledgements}

\end{document}